\documentclass[12pt,draftcls,onecolumn]{IEEEtran}

\usepackage{cite}
\usepackage{amsmath,epsfig,amssymb}
\usepackage[ruled,vlined]{algorithm2e}
\usepackage{psfrag}
\usepackage{subfigure}
  \usepackage{float}
  \usepackage{cases}
\usepackage[amsmath,thmmarks]{ntheorem}
\usepackage{color}
\usepackage{tikz}
\usepackage{graphicx}

\usepackage{booktabs,caption}
\usepackage[flushleft]{threeparttable}

\usepackage{verbatim} 
\usepackage{glossaries}

\usepackage{stfloats, flushend}

 \floatstyle{ruled}
 \newfloat{algorithm}{htb}{loa}
 \floatname{algorithm}{\textbf{Algorithm}}
 \makeatletter
 \providecommand*{\toclevel@algorithm}{0} 
 \makeatother

\theoremsymbol{$\Box$}
\newtheorem{remark}{Remark}

\theoremsymbol{$\Box$}
\newtheorem{definition}{Definition}

\theoremsymbol{$\Box$}
\newtheorem{lemma}{Lemma}

\theoremsymbol{$\Box$}

\theoremsymbol{$\Box$}
\newtheorem{proposition}{Proposition}

\theoremsymbol{$\Box$}

\theoremsymbol{$\Box$}

\theoremsymbol{$\Box$}

\theoremsymbol{$\Box$}

\makeatletter
\newcommand\unmarkfootnote[1]{%
  \begingroup
    \let\@makefntext\noindent
    \footnotetext{#1}%
  \endgroup}
\makeatother

\makeatletter

\newcommand{\Rmnum}[1]{\expandafter\@slowromancap\romannumeral #1@}
\makeatother

\newcommand{\mat}{\boldsymbol}

\usepackage{footnote}

\usepackage{xspace}
\usepackage{setspace}

\newcommand{\diag}{\mathop{\mathrm{diag}}}

\makeglossaries

\newacronym{mac}{MAC}{multiple-access channel}
\newacronym{bc}{BC}{broadcast channel}
\newacronym{mimo}{MIMO}{multiple-input multiple-output}
\newacronym{siso}{SISO}{single-input single-output}
\newacronym{af}{AF}{amplify-and-forward}
\newacronym{df}{DF}{decode-and-forward}
\newacronym{cf}{CF}{compress-and-forward}
\newacronym{mwrc}{MWRC}{multi-way relay channel}
\newacronym{pde}{PDE}{partial data exchange}
\newacronym{fde}{FDE}{full data exchange}
\newacronym{iid}{i.i.d.}{independent and identically distributed}
\newacronym{awgn}{AWGN}{additive white Gaussian noise}
\newacronym{awg}{AWG}{additive white Gaussian}
\newacronym{sic}{SIC}{successive interference cancellation}
\newacronym{snr}{SNR}{signal-to-noise ratio}
\newacronym{sinr}{SINR}{signal to interference plus noise ratio}
\newacronym{zf}{ZF}{zero-forcing}
\newacronym{mmse}{MMSE}{minimum mean square error}
\newacronym{sud}{SUD}{single user decoding}
\newacronym{dof}{DoF}{degrees of freedom}
\newacronym{gdof}{GDoF}{generalized degrees of freedom}
\newacronym{nnc}{NNC}{noisy network coding}
\newacronym{dmn}{DMN}{discrete memoryless network}
\newacronym{csi}{CSI}{channel state information}
\newacronym{ee}{EE}{energy efficiency}
\newacronym{ian}{IAN}{treating interference as noise}
\newacronym{snd}{SND}{simultaneous non-unique decoding}
\newacronym{brd}{BRD}{best response dynamics}
\newacronym{br}{BR}{best response}
\newacronym{ne}{NE}{Nash equilibrium}
\newacronym{lhs}{LHS}{left-hand side}
\newacronym{rhs}{RHS}{right-hand side}
\newacronym{ia}{IA}{interference alignment}

\begin{document}

%
\title{Semi-dynamic Green Resource Management in Downlink Heterogeneous Networks by Group Sparse Power Control}
\author{Pan Cao, 
       Wenjia Liu, 
       John S. Thompson, 
       Chenyang Yang, \\ and Eduard A. Jorswieck 
\thanks{P. Cao and J. Thompson are with the Institute for Digital Communications (IDCoM), The University of Edinburgh, Edinburgh EH3 9JL, United Kingdom (email: \{p.cao,  john.thompson\}@ed.ac.uk). 

W. Liu and C. Yang are with the School of Electronics and Information Engineering, Beihang University, Beijing 100191, China (email: \{liuwenjia, cyyang\}@buaa.edu.cn). 

E. Jorswieck is with the Chair of Communications Theory, Communications Laboratory, TU Dresden, Dresden 01062, Germany (email: eduard.jorswieck@tu-dresden.de). 
}
}

\maketitle

\begin{abstract}
This paper addresses the energy-saving problem for the downlink of heterogeneous networks, which aims at minimizing the total base stations (BSs) power consumption while each user's rate requirement is supported. The basic idea of this work is to make use of the flexibility and  scalability of the system such that more benefits can be gained by efficient resource management. This motivates us to propose a flexible BS power consumption model, which can control system resources, such as antennas, frequency carriers and transmit power allocation in an energy efficient manner rather than the "on/off" binary sleep mode for BSs. To denote these power-saving modes, we employ the group sparsity of the transmit power vector instead of the $\{0, 1\}$ variables. Based on this power model, a semi-dynamic green resource management mechanism is proposed, which can jointly solve a series of resource management problems, including BS association, frequency carriers (FCs) assignment, and the transmit power allocation, by group sparse power control based on the large scale fading values. In particular, the successive convex approximation (SCA)-based algorithm is applied to solve a stationary solution to the original non-convex problem.  Simulation results also verify the proposed BS power model and the green resource management mechanism.   

\end{abstract}


\IEEEpeerreviewmaketitle

\section{Introduction}\label{sec:intro}


The definition of the next generation (5G) networks gives the main focus on providing ubiquitous and high data rate services for massive devices: for example, {data rates of several tens of Mbit/s should be supported for tens of thousands of users}
\cite{NGMN5G}. 
To realize this 5G vision, the future 5G networks should be planned and deployed based on the \emph{peak traffic load} in an area such that all the quality of service (QoS) levels throughout the entire networks can be always satisfied. 
Networks densification and offloading
, increasing bandwidth (e.g., by spectrum sharing\cite{SpectrumSharingJorswieck2014} and carrier aggregation \cite{CarrierALTE}),
and the advanced MIMO (e.g., scaling up the number of antennas \cite{ScalingUpMIMO}) are recognized as the three key technologies in 5G networks to increase the spectral efficiency \cite{5GJSAC2014}. By employing these concepts, future 5G networks are more likely to become increasingly dense, massive and heterogeneous.
However, like a double-edge sword, these \emph{dense}, \emph{massive} and \emph{heterogeneous} advances in return may limit the network performance and increase the energy consumption if the proper resource management is not adopted. 
Therefore, if a heterogeneous network (HetNet){\footnote{Hereafter, we use the general "HetNet" to denote all the types of (single-tier or multi-tier) multi-cell environment, because our proposed mechanism is independent of the BSs' tiers/density, and the number of BSs' antennas.}} are already planned or deployed in a typical area, a question arises:
\begin{itemize}
\item[\emph{Q:}] \emph{How to make a HetNet green by resource management in operation, especially for the partially loaded scenarios?}
\end{itemize}


This question on green resource management has attracted intensive research. A brief, comprehensive, yet non-exhaustive review of related work is given as follows.

\subsection{Related Works}\label{sec:relatedwork}



The general base station (BS) and user equipment (UE) association is a popular way to improve the overall network performance by scheduling the connections between BSs and UEs such that the inter-BS interference can be properly managed, see  
\cite{BSUEass2,BSUE4} and the references therein for the HetNets.
When the green communications is the goal, a adaptive BS-UE association can be used to reduce the network energy consumption by power control. In \cite{BSminimizingPower1}, 
 both the power allocation and BS assignment in nonorthogonal downlink transmission code-division multiple-access (CDMA) communication systems are jointly studied, where each UE is allowed to connect to more than one BS.
The authors in \cite{BSpowerminimizing2} propose a joint BS association and power control algorithm to simultaneously maximize the system revenue and minimize the total transmit power consumption such that each UE can be served by the right BS. 
Two types of BS-UE association problems are addressed in\cite{BSpowerminimizingLocation} by minimizing the total network power consumption (global throughtput) and minimizing each user's power consumption (UE equilibrium), respectively.
In \cite{DecentralizedBSpowerminimizing}, BS association and downlink beamforming is jointly optimized by minimizing the sum power consumption while guaranteeing a minimum signal-to-interference-and-noise-ratio (SINR) per UE. 
Instead of studying the BS-UE association under universal spectrum reuse, a joint design of flexible spectrum assignment and BS-UE association might further improve the network performance \cite{JointBSUEFC1}. 
Another special case of spectrum reuse is orthogonal frequency division multiple access (OFDMA), which leads to a joint subcarrier assignment and BS-UE association problem. 
Some recent works on energy efficiency maximization for the downlink multi-cell OFDMA system have been addressed in \cite{EEOFDMAlargenatenna,EEschedualingOFDM,EEOFDMA2} and the references therein. 
 
In addition to the green scheduling and power allocation in the above works, another important way to save network energy consumption is to completely or partially turn off some "free" BSs with no/low load, e.g., \cite{BSsleeping1,BSsleep1,BSsleepsurvey,BSsleepingMM,5GEnergyEfficiency2014,EEOFDMAMM} and the references therein. For instance, the authors in \cite{BSsleepingMM, 5GEnergyEfficiency2014, EEOFDMAMM,SEEEyang} introduce and optimize a \{0,1\} matrix to control the "on/off" status of the BSs, and in particular  \cite{EEOFDMAMM,SEEEyang} also consider the scheduling and transmit power minimization. However, the "on/off" two-status decision might be crude and coarse, since this power model implies that all the "on" BSs consume the same constant circuit power in spite of their different traffic loads, which is not true in practical systems. This motivates that the hardware components of the networks should be as flexible and reconfigurable as possible, since this hardware flexibility and scalability can be exploited to improve energy efficiency/saving, by reconfiguring the BS components according to the effectively used resources \cite{BSsleep1,FlexiblePower}. Thus, flexibly turning off or deactivating some hardware components are preferred, e.g., the antenna muting/adaptation \cite{antennamuting,antennaswitching1}. In the time domain,  the discontinuous transmission (DTX) \cite{DTX1} based on the varying channel quality is another example of BS sleep, which is extended in \cite{BSPowerminimizing1} by combining the scheduling and power control to minimize the BS energy consumption. By adopting the BS sleep mode mechanisms, some unnecessary energy consumptions, for example, static power and part of load-dependent power for the partial-loaded BSs, can be saved.

However, the systems in most previous works on green HetNets are not as flexible and scalable as possible and are usually based on some of the following assumptions: \emph{R1. both BSs and UEs are equipped with a single antenna; R2. each BS is allowed to serve one UE at a time on each FC; R3. each UE is allowed to be connected to only one BS at a time; R4. each UE is allowed to operate on only one FC at a time; R5. each FC is not allowed to be reused by two or multiple UEs at a time; R6. simple transmit power control for each UE on a FC is adopted, e.g., fixed power allocation or fractional power control; R7. the "on/off" two-status BS sleep mode is used.} In fact, these "restricted"  system assumptions should be and can be relaxed due to recent hardware and signal processing capabilities in order to further improve the green performance.

\subsection{Contributions} \label{sec:introcontribution}

With this respect, this work is aimed to develop a power model of the HetNets involving the hardware flexibility and reconfiguration and to provide a semi-dynamic green resource management mechanism to adjust the networks energy consumption to the varying data traffic load. Inspired by the centralized benefits in the cloud technologies \cite{Cloud5G}, we assume that all the BSs in one HetNet are connected to a central processor (CP){\footnote{The CP could be either the central data center in the Cloud radio access network (C-RAN) or a macro BS who has the capability to do central optimization for the  entire network.}} via backhaul links (in fact, this work requires only a low backhaul overhead) such that a central optimization can be implemented. The idea herein is to throw away the concept or limits of the "cell" such that a more flexible association/access between BSs,  UEs, frequency carriers (FCs) is allowed under the following system assumptions: 
\begin{itemize}
\item[\emph{A1}.] \emph{Multi-Antenna System:} Each BS is equipped with multiple or even a large scale antenna array;
\item [\emph{A2}.] \emph{Dual Multi-Connectivity/Access Enabled Operation:} Each BS can simultaneously serve more than one UEs on each individual FC (i.e., multi-user transmission). Meanwhile, each UE can be simultaneously served by more than one BSs on each individual FC (e.g., the coordinated multi-point (CoMP) transmission);
\item [\emph{A3}.] \emph{Dual Multi-Carrier Enabled Operation:} Each BS and each UE can operate simultaneously on one or more FCs (e.g., through carrier aggregation  \cite{CarrierALTE});
\item [\emph{A4}.] \emph{Spectrum Reuse or Not:} Each FC is allowed to be reused by any BS set and UE set;
\item [\emph{A5}.]  \emph{Frequency-Selective Fading Channel Model:} The same communication link on different FCs may experience different channel qualities; 
\item [\emph{A6}.]  \emph{Flexible Transmit Power Allocation:} Flexible downlink transmit power is allocated subject to the per-BS transmit power budget. 
\end{itemize}
These general system operation assumptions allow us to formulate a series of the flexible scheduling and efficient resource management problems: such as \emph{P1. BS-UE association problem (BS-selection and ''many-to-many" assignment)}, \emph{P2. BS/UE-FC assignment problem (FC-selection and ''many-to-many" assignment)}, \emph{P3. downlink transmit power allocation problem}, \emph{P4. intra-carrier interference management problem (a side-product of P1-P3)}, and \emph{P5. flexible BS power model (multiple sleeping modes enabled)}.
In order to jointly solve the above resource management problems, we consider all the BSs, FCs, time blocks, transmit power as the "resources" in the HetNet, and pour them into the "pool" (i.e., the CP). The output of a predefined central optimization of green resource management based on a flexible and scalable power consumption model will give the answer to Question \emph{Q}. 

The main contributions along with the organization of this paper are listed as follows.

\begin{itemize}

\item In Section \ref{sec:powermodel},  inspired by \cite{GreengroupsparseBeamforming,BSUE9}, we employ for the first time the $\ell_0$ norm of the power vector in place of the $\{0, 1\}$ matrix to control the "on/off"of hardware components according to the effectively assigned FCs. With this choice, BSs' signal processing power can be flexibly scaled by group sparse power control. Based on this idea, a flexible and scalable BS downlink power consumption model is proposed.

\item 
In Section \ref{sec:problemformulation}, we formulate a \emph{semi-dynamic} downlink network energy consumption minimization problem using slowly varying the large scale fading (LSF) values. This problem enables us to jointly optimize all the green BS-UE association and FC assignment, transmit power allocation, BS deactivation.
Since this problem is shown to be a NP-hard problem, we apply a successive convex approximation (SCA)-based algorithm in Section \ref{sec:algorithm} to solve it efficiently, and its convergence to a stationary solution is proved. 


\end{itemize}

\emph{Notations:} $|\mathcal{X}|$ and $|\mat{x}|$ denotes the number of the elements of a set $\mathcal{X}$ and a vector $\mat{x}$; $\mathcal{X}(i)$ denotes the $i$-th element in the set $\mathcal{X}$; $\mathcal{X}_1\backslash\mathcal{X}_2$ denotes the set $\mathcal{X}_1$ but excluding all the elements in the set $\mathcal{X}_2$; $\diag[\mat{x}]$ denotes a diagonal matrix with the elements in $\mat{x}$ as its diagonal elements; $\binom nL$ denotes the number of $n$-combinations for a $L$-element set.

\section{System Model} \label{se: SM}

Consider the downlink transmission in one HetNet, where $K$ BSs communicate with $L$ active single-antenna UEs employing $F$ orthogonal FCs. Let $\mathcal{K} \triangleq \{1, 2, \cdots, K\}$, $\mathcal{L} \triangleq \{1, 2, \cdots, L\}$ and $\mathcal{F} \triangleq \{1, 2, \cdots, F\}$ denote the index set of the BSs, UEs and FCs, respectively. This setup is denoted by $\mathcal{K} \times \mathcal{L} \times \mathcal{F}$.
Based on the general system assumptions \emph{A1-A6} in Section \ref{sec:introcontribution}, we let $N_k$ and  $W_f$ Hz denote the number of antenna of BS $k\in\mathcal{K}$ and the bandwidth of FC $f \in \mathcal{F}$. Let $p_{k,\ell}^f\geq 0$ denote the downlink transmit power at BS $k$ allocated for the transmission to UE $\ell  \in \mathcal{L}$ on FC $f\in\mathcal{F}$. The transmit power $\{p_{k,\ell}^f\}_{\ell\in\mathcal{L}, f\in\mathcal{F}}$ at each BS $k$ are allowed to be flexibly allocated to the $LF$ channels but subject to the per-BS transmit power budget $P_{BS,k}^{max}$, i.e., $\sum_{\ell=1}^L\sum_{f=1}^Fp_{k,\ell}^f \leq P_{BS,k}^{max}$. 

\subsection{Channel Model}

We assume that the channel on each FC is quasi-static block-fading which is constant for a number of \emph{symbol intervals}.\footnote{The symbol interval denotes the time consumed for a transmission of one symbol.} A symbol interval is denoted by $T_{ST}$.
Let $\mat{h}_{k,\ell}^{f} =\sqrt{\alpha_{k,\ell}^{f}}\tilde{\mat{h}}_{k,\ell}^{f}  \in \mathbb{C}^{N_k \times 1}$ be the instantaneous channel state information (CSI) from BS $k \in \mathcal{K}$ to UE $\ell \in \mathcal{L}$ on FC $f \in \mathcal{F}$ in a certain time slot, where $\alpha_{k,\ell}^{f}$ denotes the LSF gain including path loss and shadowing, and $\tilde{\mat{h}}_{k,\ell}^{f}$ denotes the corresponding small scale fading (SSF) vector where each entry is assumed to satisfy independent and identically distribution (i.i.d.)  with zero mean and unit covariance.


The \emph{age of LSF (A-LSF)} is defined as the time duration over which the LSF of a communication link is considered to be not varying. The time duration over which the SSF stays constant is in fact the \emph{coherence time}. 
In many mobile radio situations, the A-LSF is usually tens or hundreds of times longer than the coherence time.
Since different carrier frequencies result in different length of the coherence time,
we let $T^f_{SSF}$ denote the coherence time on FC $f\in\mathcal{F}$. 
 Without loss of generality, we assume the channels on different FCs have the same A-LSF (denoted by $T_{LSF}$) and coherence bandwidth, and then
about $\beta_{1,f} = \lfloor\frac{{T}_{LSF}}{{T}_{ST}}\rfloor$ and $\beta_{2,f} = \lfloor\frac{{T}^f_{SSF}}{{T}_{ST}}\rfloor$ symbols can be transmitted for the fixed LSF and the fixed channel on FC $f$, respectively. 

\subsection{Resource Management Mechanism}


In terms of resource management, the dynamic design based on the instantaneous CSI significantly gains the benefits by adjusting strategies with the varying CSI but at the  cost of high complexity. In most practical mobile communication scenarios,  it is usually not allowed to design complicated instantaneous transmission strategies (e.g., by the high overhead required and high-complexity iterative algorithms) because of the limited coherence time. In contrast, the long-term fixed transmission strategies for a long time duration has a very low complexity but usually results in a very inefficient usage of the resources because of the mismatch between the fixed strategies and the varying CSI. 
This motivates us to design a \emph{semi-dynamic} hybrid resource management mechanism:
\begin{itemize}
\item [\emph{M1}.] \emph{Maximum Ratio Transmission (MRT) Beamforming:} During each coherence time, the low-overhead and low-complexity MRT downlink beamforming scheme is used. Each BS can design the MRT beamforming patterns for its serving UEs \emph{independently} and \emph{locally} based on only the instantaneous CSI of the desired links, which has a low computation time (the remaining time can be left for uplink/downlink transmission) and no backhaul overhead needed for the SSF.{\footnote{This is also the motivation for us not to use the joint precessing but to use independent signals enhancement in the CoMP.}} One beamforming design is sufficient for the whole coherence time;   
\item  [\emph{M2}.] \emph{Resource Management:} During each A-LSF, green resource management problem is optimized at the CP based on only the LSF values. Only one implementation is needed for the whole A-LSF hence we call it "semi-dynamic".
\end{itemize}

In \emph{M1}, no optimization but only the computation of the simple MRT beamforming pattern is required. Our main focus will be on the optimization in \emph{M2}, which only requires that the LSF values are available at the CP. 

\subsection{Channel Acquisition}
In order to implement \emph{M1} and \emph{M2}, the acquisition of SSF and LSF are required, respectively. Some symbol intervals within each coherence time might be taken for channel training, e.g., by the pilot sequence transmission, and the remainder is left for downlink data symbol transmission{\footnote{The uplink data transmission is not considered here in order to focus on the downlink transmission, since the total network energy is mainly consumed by BSs in the downlink transmission. Otherwise, it is equivalent to consider the "coherence time" used in this work to be a shorter one excluding the uplink transmission time.}}.

In this work, the time-division duplexing (TDD) operation scheme is employed, because the feedback phase under the frequency-division duplexing (FDD) operation can be eliminated by using the \emph{channel reciprocity} and additionally the pilot overhead might be reduced for the multi-antenna system, especially for the massive MIMO system. In the uplink channel training,  all UEs transmit pilot sequences to their associated BSs on the assigned FCs. 
Let $\sqrt{\tau_f}\mat{\phi}_{\ell}^f$ with $|| \mat{\phi}_{\ell}^f||=1$ be the training vector with the length $\tau_f$ transmitted from UE $\ell$ with the transmit power $p_{UE,\ell}^f$ to its associated BS $k$ on an assigned FC $f$. 

Let $\mathcal{U}_{FC,f} \subseteq \mathcal{L}$ denote the set of UEs who reuse FC $f$.\footnote{In fact, the channel estimation is done after the scheduling in \emph{M2}: BS-UE association and BS/UE-FC assignment. Thus, each UEs' set $\mathcal{U}_{FC,f}, \forall f\in \mathcal{F}$ and their serving BSs are already known before channel estimation. Without loss of generality, we assume $\mathcal{U}_{FC,f} \neq \emptyset$.} Then, a  $\tau_f \times |{\mathcal{U}_{FC,f}}|$ pilot sequence matrix is needed for channel training from $\mathcal{U}_{FC,f} $ to their associated BSs
\begin{align}
{\mat{\Phi}}^f = [\mat{\phi}^f_{{\mathcal{U}_{FC,f}}(1)}; \cdots; \mat{\phi}^f_{{\mathcal{U}_{FC,f}}(|{\mathcal{U}_{FC,f}}|)}].
\end{align}
If $\tau_f\geq |{\mathcal{U}_{FC,f}}|$, we can generate the pairwise orthogonal pilot sequences $\{\mat{\phi}^f_{\ell}\}_{\ell \in\mathcal{U}_{FC,f}}$. Otherwise, pilot reuse among the UEs in ${\mathcal{U}_{FC,f}}$ is needed and pilot contamination exists. To consider the both cases, we generally denote by $\mathcal{U}_{FC,f}^{m} \subset \mathcal{U}_{FC,f}$ the set of UEs who use the same pilot sequence $\mat{\phi}^f_{{\mathcal{U}_{FC,f}}(m)}$ in ${\mat{\Phi}}^f$. If  $|\mathcal{U}_{FC,f}^{m}|=1$, it means no reuse of $\mat{\phi}^f_{{\mathcal{U}_{FC,f}}(m)}$. Otherwise, $|\mathcal{U}_{FC,f}^m|$ UEs reuse the same pilot sequence $\mat{\phi}^f_{{\mathcal{U}_{FC,f}}(m)}$.

When $\sqrt{\tau_f}{\mat{\Phi}^f}$ is transmitted from $\mathcal{U}_{FC,f}$ with the uplink training power
\begin{align}
{\mat{P}}_{\mathcal{U}_{FC,f}} = \diag\left[p_{UE, {\mathcal{U}_{FC,f}}(1)}^f, \cdots,  p_{UE, {\mathcal{U}_{FC,f}}(|{\mathcal{U}_{FC,f}}|)}^f\right] \in \mathbb{R}^{ |{\mathcal{U}_{FC,f}}| \times |{\mathcal{U}_{FC,f}}|},
\end{align}
the $\tau_f \times N_k$ received signal at BS $k$ on FC $f$ can be expressed as
\begin{align}
\mat{Y}_k^f = \sqrt{\tau_f}\mat{P}_{\mathcal{U}_{FC,f}}^{\frac{1}{2}}\mat{\Phi}^f\mat{A}_{k}^{f,\frac{1}{2}}\tilde{\mat{H}}_{k}^f + \mat{Z}_{k}^f,
\end{align}
where 
\begin{align}
&\tilde{\mat{H}}_k^f = [\tilde{\mat{h}}_{k,\mathcal{U}_{{FC,f}(1)}}^{f,T}; \cdots; \tilde{\mat{h}}_{k,\mathcal{U}_{FC,f}(|{\mathcal{U}_{FC,f}}|)}^{f,T}] \in \mathbb{C}^{ |{\mathcal{U}_{FC,f}}| \times N_k}, \\
&{\mat{A}}_k^f = \diag\left[\sqrt{\alpha_{k,\mathcal{U}_{FC,f}(1)}^f}, \cdots, \sqrt{\alpha_{k,\mathcal{U}_{FC,f}(|\mathcal{U}_{FC,f}|)}^f}\right] \in \mathbb{R}^{ |\mathcal{U}_{FC,f}| \times |{\mathcal{U}_{FC,f}}|}, \\
&{\mat{Z}}_k^f = [\mat{z}_{k,1}^f, \cdots, \mat{z}_{k, N_k}^f] \in \mathbb{C}^{\tau \times N_k},
\end{align}
where $\mat{z}_{k,n}^f \in \mathbb{C}^{\tau_f\times 1}, \forall n\in \{1, \cdots, N_k\}$ denotes the noise vector at $n$-th antenna of BS $k$ in uplink training phase on FC $f$. We assume that $\mat{z}_{k,n}^f \sim\mathcal{C}\mathcal{N}(\mat{0},W_f\sigma^2\mat{I}),~\forall n$, since the terminal noise linearly increases with the carrier bandwidth.

\begin{lemma}
The minimum mean square error (MMSE) estimate of the channel from a typical UE $\ell \in \mathcal{U}_{FC,f}$ to its associated BS $k$ on FC $f$ is $\sqrt{\alpha_{k,\ell}^{f}}\hat{\mat{h}}_{k,\ell}^f $, where 
\begin{align}
\hat{\mat{h}}_{k,\ell}^f = \frac{\sqrt{\tau_f\alpha_{k,\ell}^fp_{UE,\ell}^f}}{\tau_f\alpha_{k,\ell}^fp_{UE,\ell}^f +  { \sum_{j\in\mathcal{U}_{FC,f}^{\ell}\backslash\{\ell\}}\tau_f\alpha_{k,j}^fp_{UE,j}^f } + W_f\sigma^2} \mat{\phi}_{\ell}^{f,H}\mat{Y}_k^f.
\end{align}
The SSF $\tilde{\mat{h}}_{k,\ell}^{f}$ can be expressed as 
\begin{align}
\tilde{\mat{h}}_{k,\ell}^{f} = \hat{\mat{h}}_{k,\ell}^{f}  + \mat{e}_{k,\ell}^{f},  \label{eq:channelesti}
\end{align}
where $\hat{\mat{h}}_{k,\ell}^{f} \sim \mathcal{C}\mathcal{N}(\mat{0}, \delta_{k,\ell}^f\mat{I})$ is independent of the estimation error ${\mat{e}}_{k,\ell}^{f}\sim \mathcal{C}\mathcal{N}(\mat{0},(1-\delta_{k,\ell}^f) \mat{I})$ with
\begin{align}
\delta_{k,\ell}^f \triangleq \frac{\tau_f p_{UE,\ell}^f \alpha_{k,\ell}^f}{\tau_f p_{UE,\ell}^f \alpha_{k,\ell}^f+  \sum_{j\in\mathcal{U}_{FC,f}^{\ell}\backslash\{\ell\}}\tau_f\alpha_{k,j}^fp_{UE,j}^f  + W_f\sigma^2}. \label{eq:channelerror2}
\end{align} 
\end{lemma}

\begin{IEEEproof}
This result is following the standard MMSE estimation in \cite[Chapter 15.8]{MMSEestimation}.
\end{IEEEproof}
\begin{remark}
When no pilot sequence is reused, the channel estimation error might become negligible (i.e., $\delta_{k,\ell}^f \rightarrow 1$) if $\tau_f p_{UE,\ell}^f\alpha_{k,\ell}^f$ is sufficiently large and $W_f$ is not very large. Interestingly, \eqref{eq:channelerror2} also implies that pilot sequences can be reused on the same FC without significant performance loss by those UEs as long as they have low LSF gains or low uplink training power to the same BS.
In addition, the assignment of FCs and their bandwidth also influences the channel estimation $\delta_{k,\ell}^f$ in \eqref{eq:channelerror2}, since the same link experiences different LSF on different FCs. 
\end{remark}

In terms of the LSF,  it depends on the communication environment and mainly on the geo-locations of the UEs because of the path-loss. This motivates us to employ the LSF map \cite{LSFMapLi} to denote the LSF of different geo-locations.

\begin{definition}
A LSF map is defined as the set of LSF values of the dense sampling geo-locations in a geographic area.
A "point" on the LSF map contains $KF$-dimensional LSF values of the downlink channels{\footnote{A LSF map probably also contains the LSF values of the uplink channels for the FDD system.}} from $K$ BSs to the corresponding geo-location on the $F$ FCs, respectively. 
\end{definition} 

A LSF map can be generated offline by measuring the LSF values of the sampling geo-locations in advance once the deployment of BSs is given \cite{ChannelMapMeasurement}, and thus it can be used as a \emph{prior information} (stored at the BS or the CP) to perform the cooperative semi-dynamic resource allocation in \emph{M2}.
For example, combining the LSF map and current UEs' geo-locations (maybe provided by GPS), the LSF values in the next A-LSF can be determined based on the UEs' mobility prediction \cite{MobilityPrediction}.



\subsection{Initial BS-UE Association}

Let $\mathcal{U}_{k}^f \subseteq \mathcal{L}$ and $\mathcal{B}_{\ell}^f \subseteq \mathcal{K}$ denote the UEs set simultaneously served by BS $k \in \mathcal{K}$ and the BSs set simultaneously serving UE $\ell \in \mathcal{L}$, respectively, on FC $f\in\mathcal{F}$. Note that some UEs in $\mathcal{U}_{k}^f$ might not be in the "cell" of BS $k$ because of the CoMP transmission. 

\begin{lemma}
For the setup $\mathcal{K}\times \mathcal{L}\times \mathcal{F}$, there exist at most $\sum_{k=1}^K\sum_{n=1}^{\min(FN_k, L)} \binom nL$ possible solutions to the BS-UE association problem in {P1}.
\end{lemma}
\begin{IEEEproof}
In principle, it is possible for each BS $k$ equipped with $N_k$ antennas to simultaneously and independently serve up to $N_k$ UEs on each FC $f$, and thus up to $\min(FN_k, L)$ UEs can be served by BS $k$ if it serves different UEs set on different FCs (i.e., $\mathcal{U}_{k}^f \cap \mathcal{U}_{k}^{\overline{f}}  = \emptyset, \forall f \neq \overline{f}$). Then, the proposed result can be obtained by solving a combinatorial problem.
\end{IEEEproof}



In order to remove unlikely solutions to reduce the complexity, we propose an initial BS-UE association to shrink the solutions set as follows.
Each BS $k$ with $N_k$ antennas initially selects $N_k$ UEs with the strongest LSF gains on each FC, based on the LSF map and the UEs' mobility prediction, to form its initial UEs set. Since the LSF mainly depends on the UEs' geo-locations, the $N_k$ UEs with the strongest LSF gains generally are the closest $N_k$ UEs to BS $k$. Therefore, we have $\mathcal{U}_k \triangleq \mathcal{U}_k^1=\mathcal{U}_k^2=\cdots \mathcal{U}_k^F$ where $|\mathcal{U}_k|=N_k$.
After selecting UEs by all BSs, each UE $\ell \in \mathcal{L}$ might be simultaneously selected by multiple BSs for the potential CoMP transmission. We let $\mathcal{B}_{\ell} \triangleq \mathcal{B}_{\ell}^1=\mathcal{B}_{\ell}^2=\cdots \mathcal{B}_{\ell}^F$ denote the initial BSs set consisting of all the serving BSs who initially select UE $\ell$. 
\begin{remark}
In general, it is reasonable to assume that each UE $\ell \in \mathcal{L}$ is initially selected by at least one BS, i.e., $|\mathcal{B}_{\ell}| \geq1$. In fact, it is rare that a UE cannot be initially selected by any BS, since the BSs are equipped with multiple antennas and the BS deployment is in practical based on UEs' distribution and density. If it really happens, it means that there exist more UEs than the network capacity or the non-selected UEs suffer from very bad channel conditions, and thus they should be deactivated during the next A-LSF. In this case, the proposed initial BS-UE association scheme also includes a simple user selection scheme. 
 \end{remark}

After the initial BS-UE association, the number of feasible solutions to Problem \emph{P1} is reduced to $\Pi_{\ell=1}^L \left(|\mathcal{B}_{\ell}|!\right)$, thereby resulting in $\Pi_{\ell=1}^L \left(|\mathcal{B}_{\ell}|!\times F!\right)$ feasible solutions to the BS/UE-FC assignment problem \emph{P2}. 
The optimal scheduling solution can by further determined in the resource management \emph{M2} by power control.

\section{BSs Energy Consumption Model} \label{sec:powermodel}

For the setup $\mathcal{K}\times \mathcal{L} \times \mathcal{F}$ after initial BS-UE association, the downlink transmit power $\{p_{k,\ell}^f\}_{k\in\mathcal{B}_{\ell},\ell\in\mathcal{L},f \in \mathcal{F}}$ forms an \emph{irregular}{\footnote{The irregularity is because $|\mathcal{B}_{\ell}|$ might be different for each UE $\ell$.}} three-dimensional "tensor" with the size of $ |\mathcal{B}_{\ell}| \times L \times F$. In particular, the status of a link from BS $k$ to UE $\ell$ on FC $f$ can be implied by $p_{k,\ell}^f$. More precisely, the link is \emph{on} if $p_{k,\ell}^f > 0$. Otherwise, it is \emph{off}. This motivates us to propose a general BSs downlink energy consumption model based on the transmit power control. 

\subsection{BSs Downlink Energy Consumption Model}

Before showing the energy consumption model, we first give some definitions.
\begin{definition}\label{de:BSpower}
\begin{itemize}

\item Let $\mat{p}_{BS, k}^f \triangleq \left[ {p}_{k, \mathcal{U}_{k}(1)}^{f}, {p}_{k, \mathcal{U}_{k}(2)}^{f}, \cdots, {p}_{k, \mathcal{U}_{k}(| \mathcal{U}_{k}|)}^{f}\right]^T  \in \mathbb{R}_+^{|\mathcal{U}_{k}|\times 1}$ denote the transmit power of BS $k$ to all the UEs in $\mathcal{U}_{k}$ on FC $f$; 

\item Let $\mat{p}_{BS, k} \triangleq \left[ \mat{p}_{BS, k}^1 , \mat{p}_{BS, k}^2, \cdots, \mat{p}_{BS, k}^F \right]^T  \in \mathbb{R}_+^{|\mathcal{U}_{k}|F\times 1}$ denote the transmit power of BS $k$ to all the UEs in $\mathcal{U}_{k}$ on all the FCs; 

\item Let $ \mat{p} \triangleq [\mat{p}_{BS,1}, \mat{p}_{BS,2}, \cdots, \mat{p}_{BS, K}]^T \in \mathbb{R}_+^{F\sum_{k=1}^K|\mathcal{U}_k|\times 1}$ denote the transmit power at all the $K$ BSs to their all initially selected UEs on all the FCs.
 \end{itemize}
\end{definition}

Let $\mat{T}_{BS,k}$ and $\mat{T}_{BS,k}^f$ denote $F|\mathcal{U}_{k}| \times  F\sum_{k=1}^K|\mathcal{U}_k|$ and $|\mathcal{U}_{k}| \times  F\sum_{k=1}^K|\mathcal{U}_k|$ selective matrices only consisting of $\{0, 1\}$ such that $\mat{p}_{BS, k} = \mat{T}_{BS,k} \mat{p}$ and $\mat{p}_{BS, k}^f = \mat{T}_{BS,k}^f \mat{p}$, respectively.

In the initial BS-UE association, each BS $k$ is allowed to connect to $N_k$ UEs on all $F$ FCs. However, this \emph{maximum-connectivity}
scenario rarely happens because it is usually neither necessary nor optimal to achieve certain UEs' transmission rate requirement because of the limits of intra-carrier interference and per-BS power constraint. 
Therefore, many elements of $\mat{p}_{BS,k}$ and $\mat{p}$ 
would be zeros, which implies that these transmit power vectors have the (group) sparse property. For example, BS $k$ will be in \emph{deep-sleep} if $\mat{p}_{BS, k}=\mat{0}$. Otherwise, it will be \emph{active}.
Inspired by this sparsity property, we propose to employ the \emph{group sparsity} of the transmit power vectors to illustrate the status of the  BSs or FCs.



\begin{definition}\label{de:groupsparsity}
A vector is group sparse if it has a grouping of its components and the components within each group are likely to be either all zeros or not.
Let $\mat{x}\triangleq [\mat{x}_1, \mat{x}_2, \cdots, \mat{x}_G]$ be a $M\times 1$ vector with $G$ non-overlapping groups, where the vector $\mat{x}_g$ denotes the $g$-th group of the size $M_g \times 1$ satisfying $\sum_{g=1}^GM_g = M$. The weighted group sparsity of the vector $\mat{x}$ is defined by
\begin{align}
||\mat{x}||_{0,\mat{w}}^{G, M_g} \triangleq \sum_{g=1}^G w_g \cdot \mathrm{sign}(|| \mat{x}_g||_0), \label{eq:wgs}
\end{align}
where $\mat{w} \triangleq [w_1, w_2, \cdots, w_G]$ with $w_g$ as the weight of the group $\mat{x}_g$ and
 \begin{subnumcases}
{\mathrm{sign}(|| \mat{x}_g||_0) =} 
 0 & $\mathrm{when}~ \mat{x}_g=\mat{0}$   \label{eq:lineaConeq}\\  
 1 & $\mathrm{otherwise}$. \label{eq:lineaCongeq}
\end{subnumcases}
When $\mat{w}=\mat{1}$, we use $||\mat{x}||_{0}^{G, M_g}$ to denote the standard unweighted group sparsity $\ell_0$ norm.


\end{definition}

For example, $||\mat{p}||_{0}^{K, F|\mathcal{U}_k|}$ can be used to count the number of active BSs. 
Therefore, we propose to employ the group sparsity of the transmit power vectors to model the downlink BSs energy consumption.


\begin{proposition}
The BSs power consumption model can be assumed as
\begin{align}
P_{BS} \triangleq \underbrace{\sum_{k=1}^K P_k^{sleep_0}
 + \sum_{k=1}^K || \mat{p}_{BS,k}||_{0,\mat{\mu}_k}^{F, |\mathcal{U}_{k}|}  }_{circuit ~\& ~signal ~processing ~power}   
&+ \underbrace{\sum_{f=1}^F\left(1-\frac{\tau_f}{\beta_{2,f}}\right)\sum_{k=1}^K {\frac{1}{\eta_k}\mat{1}^T \mat{p}_{BS,k}^f}}_{downlink~transmit ~power}
+   \underbrace{P_{haul}\frac{R_{haul}}{100 Mbits/s}}_{backhaul~power}
  \label{eq:totalpowermodel}
\end{align}
where $P_k^{sleep_0}$ denotes the basic static power consumption to support the "deep-sleep" mode of BS $k$; and
$\mat{\mu}_k \triangleq [P_{sp,k}^1, P_{sp,k}^2, \cdots, P_{sp,k}^F]$ denotes the weights for the weighted group sparsity
where $P_{sp,k}^f$ denotes the weight for the $f$-th group of $\mat{p}_{BS,k}$ and is expressed by \cite{ParameterBSPower}
\begin{align}
P_{sp,k}^f = N_k\frac{W_f}{10~MHz} (P_{BB}'+P_{RF}'),  \label{eq:fsppowerb}
\end{align}
where $P_{BB}'$ and $P_{RF}'$ are some reference baseband and RF related signal processing power consumption per 10 MHz bandwidth; and $\eta_k \in (0, 1)$ denotes the downlink power amplifier (PA) efficiency ratio of BS $k$;
and $P_{haul}$ is the reference backhaul power consumption for a backhaul collection of wireless links of 100 Mbit/s capacity \cite{Backhaulpower} and $R_{haul}$ is the
average total backhaul transmission rate. 
\end{proposition}


\subsection{Explanation: Terms in Power Consumption Model}

The proposed BS power consumption model in \eqref{eq:totalpowermodel}  is explained term by term as follows:

\emph{1. Circuit \& Signal Processing Power\cite{ParameterBSPower}:} 1) $\sum_{k=1}^KP_{k}^{sleep_0}$ denotes the very basic \emph{static} power consumption of all the BSs to support their "deep-sleep" mode, where $P_k^{sleep_0}$ is the "deep-sleep" power consumption of BS $k$, e.g., the power consumed by the DC-DC power supply, mains supply and active cooling system. This static power $P_{k}^{sleep_0}$ is usually different for different types of BSs. 
2) $\sum_{k=1}^K || \mat{p}_{BS,k}||_{0,\mat{\mu}_k}^{F, |\mathcal{U}_{k}|}$ denotes the power consumption by the baseband (BB) interface and the signaling of RF transceiver (RF-TRX) of all the BSs. The power consumption of the BB interface is mainly contributed by carrier aggregation, filtering, FFT/IFFT, modulation/demodulation, signal detection, channel coding/decoding, and the RF-TRX power consumption mainly depends on the bandwidth, the number of antennas and the resolution of the analogue-to-digital conversion. 
\begin{remark}
When BS $k$ is in "deep-sleep" mode, its signal processing power $\{p_{sp,k}^{f}\}_{f=1}^F$ is equal to zero. We employ $||\mat{p}_{BS,k}||_{0,\mat{\mu}_k}^{F,|\mathcal{U}_{k}|}$ to count the number of effective FCs assigned to BS $k$, which allows that each BS to have
 $(F!+1)$-level signal processing power by turning off partial hardware components according to different effective (assigned) bandwidth{\footnote{From \eqref{eq:fsppowerb}, it implies that the signal processing power for each FC is different if all individual FCs have different bandwidth.}}. This term is load-dependent. 
 For example, if a BS is required to support a high data load of UEs, more FCs should be assigned but at the cost of high signal processing power. In contrast, a BS is placed into ''deep-sleep'' if no FC is needed. Therefore, multi-level signal processing power can be perfectly determined by the group sparsity power control based on the UEs' rate requirement.  
\end{remark}

\emph{2. Downlink Transmit Power:} A BS or UE can operate \emph{simultaneously and in parallel} on different FCs (similar to the FDD mode). 
This parallel operation allows different length of pilot sequences for channel training on different FCs. The parameter $1-\frac{\tau_f}{\beta_{2,f}}$ denotes the ratio of downlink transmission time to the whole time period on a typical FC $f$. 
This term computes the total downlink transmit power consumption by all the BSs on all the FCs, while in fact, only the transmit power of the assigned FCs are counted because $\{p_{k,\ell}^f\}$ are zeros for BSs in deep-sleep and un-assigned FCs.

 \emph{3. Backhaul Power:} This term is to measure the power consumption by the backhaul overhead, usually including the exchange of the CSI, transmission data and the signaling between coordinated BSs (e.g., in the iterative processing). The backhaul power consumption highly depends on the mechanism/algorithm itself. For instance, our proposed semi-dynamic resource management mechanism has no need for the backhaul communication during the channel training and the local MRT beamforming pattern design. its main requirement is to release the downlink users data to their associated BSs. Therefore, in our scenario the average total resulting backhaul rate for each UE is approximately its average downlink data rate{\footnote{In this setup, synchronization signaling and the power allocation result releasement from the CP are also needed via backhaul links, which are not considered herein because of their very low overhead.}}, thereby
\begin{align}
R_{haul} \approx \sum_{\ell=1}^LR_{\ell}(\mat{p}),  \label{eq:backhaulrate}
\end{align} 
where $R_{\ell}(\mat{p})$ is defined in bits/s as the average downlink transmission rate for UE $\ell$.
The proposed BS energy consumption model in \eqref{eq:totalpowermodel} is expressed as a function of transmit power vector $\mat{p}$. This implies that a series of resource management problems, such as the trade-offs between the BSs energy consumption and downlink transmission rate and Problems \emph{P1-P4} in Section \ref{sec:introcontribution}, can be jointly solved by optimizing a single variable $\mat{p}$. 



\section{Downlink Transmission Rate and Problem Formulation} \label{sec:problemformulation}

In this work, we desire to minimize the BSs energy consumption while each UE's required downlink rate is guaranteed. The downlink rate of an individual UE is first derived as follows. 

 \subsection{Downlink Transmission Rate}


Given an initial BS-UE association, the average transmission rate of each UE $\ell \in \mathcal{L}$ during $T_{LSF}$ can be expressed as
\begin{align}
R_{\ell}(\mat{p}) = \sum_{f=1}^F \left(1-\frac{\tau_f}{\beta_{2,f}}\right)W_f R_{\ell}^f  \label{eq:UErate}
\end{align}
where $1-\frac{\tau_f}{\beta_{2,f}}$ denotes the downlink data transmission time fraction, and  $R_{\ell}^f$ denotes the rate contribution from $\mathcal{B}_{\ell}$ to UE $\ell$ on FC $f$, i.e.,
\begin{align} 
R_{\ell}^f &= \mathbb{E}_{\tilde{\mat{h}}}\Bigg\{\log_2\Bigg(1+\frac{\sum_{k \in \mathcal{B}_{\ell}} |\mat{h}_{k, \ell}^{f,H} \mat{w}_{k,\ell}^f|^2  }{ W_f\sigma^2 + \underbrace{\sum_{\overline{k}\in \mathcal{K}\backslash\mathcal{B}_{\ell}} \sum_{j \in \mathcal{U}_{\overline{k}}} |\mat{h}_{\overline{k}, \ell}^{f,H} \mat{w}_{\overline{k},j}^f|^2}_{\mathrm{Inter-BS_{\ell}^f}} + \underbrace{\sum_{k \in \mathcal{B}_{\ell}}\sum_{\overline{\ell} \in \mathcal{U}_{k}\backslash\{\ell\}} |\mat{h}_{k, \ell}^{f,H} \mat{w}_{k,f}^f|^2}_{\mathrm{Intra-BS_{\ell}^f}}}\Bigg)\Bigg\} \label{eq:rate1} 
\end{align} 
where $\mathbb{E}_{\tilde{\mat{h}}}\{\}$ denotes the expectation only with respect to the SSF within each $T_{LSF}$ because the LSF stays constant within $T_{LSF}$, and $\mat{w}_{k,\ell}^f \in \mathbb{C}^{N_k \times 1}$ denotes the instantaneous downlink beamforming designed based on the estimated CSI at BS $k$ for UE $\ell$ on FC $f$, and $\mathrm{Inter-BS_{\ell}^f}$ and $\mathrm{Intra-BS_{\ell}^f}$ denote the inter-BS and the intra-BS interference to UE $\ell$ on FC $f$, respectively. 


\begin{lemma}\label{le:averagerate}
By using the MRT beamforming $\mat{w}_{k,\ell}^f = \sqrt{p_{k,\ell}^f}\overrightarrow{\hat{\mat{h}}}_{k,\ell}^f$ where $p_{k,\ell}^f$ is the fixed downlink transmit power within $T_{LSF}$ and $\overrightarrow{\hat{\mat{h}}}_{k,\ell}^f \triangleq \frac{{\hat{\mat{h}}}_{k,\ell}^f}{||{\hat{\mat{h}}}_{k,\ell}^f||}$, the average rate $R_{\ell}^f$ in \eqref{eq:rate1} is approximately expressed as
\begin{align} 
R_{\ell}^f 
\approx \log_2\Bigg(1+\frac{\sum_{k \in \mathcal{B}_{\ell}} p_{k,\ell}^f\alpha_{k,\ell}^f \left(\delta_{k,\ell}^f(N_{k}-1) + 1\right)}{ W_f\sigma^2 + {\sum_{\overline{k}\in \mathcal{K}\backslash\mathcal{B}_{\ell}} \sum_{j \in \mathcal{U}_{\overline{k}}} p_{\overline{k},j}^f \alpha_{\overline{k}, \ell}^f} + {\sum_{k\in\mathcal{B}_{\ell}}\sum_{\overline{\ell} \in \mathcal{U}_{k}\backslash\{\ell\}} p_{k,\overline{\ell}}^f \alpha_{k,\ell}^f}}\Bigg), \label{eq:mrtrate}
 \end{align} 
where $\delta_{k,\ell}^f$ is defined in \eqref{eq:channelerror2}.
\end{lemma}
\begin{IEEEproof}
See Appendix \ref{subsec:ProofRate}.
\end{IEEEproof}

\begin{remark}
The approximation is because $\mathbb{E}_x\{\log_2(1+\frac{f_1(x)}{f_2(x)})\} \approx \log_2(1+\frac{ \mathbb{E}_x\{f_1(x)\}}{ \mathbb{E}_x\{f_2(x)\}})$ is used, which is widely used and partially justified in the performance analysis for the multi-antenna systems (e.g., \cite{Rateapprox2}). In particular, the simulations in \cite{MMIMOAntennas} imply this approximation has a high accuracy, especially for the large antenna array.
\end{remark}

\subsection{Problem Formulation}

A semi-dynamic green resource management problem of BSs power minimization by group sparse power control is formulated as follows
\begin{subequations} \label{eq:P1}
\begin{align}
\min_{\mat{p}\geq\mat{0}} ~~&P_{BS} \label{eq:P1a}  \\
\mathrm{s.t.}~~
&\sum_{f=1}^F \left(1-\frac{\tau_f}{\beta_{2,f}}\right)W_fR_{\ell}^f \geq {\gamma}_{\ell},~\forall \ell \in \mathcal{L}   \label{eq:P1b} \\
&\mat{1}^T(\mat{T}_{BS,k}\mat{p}) \leq {P}_{BS,k}^{max},~\forall k \in \mathcal{K}   \label{eq:P1c}
\end{align} 
\end{subequations}
where the objective function $P_{BS}$ is shown in \eqref{eq:totalpowermodel}, and the $R_{\ell}^f$ in downlink transmission rate constraint \eqref{eq:P1b} is based on \eqref{eq:mrtrate},
and the constraint \eqref{eq:P1c} denotes per-BS transmit power constraint because of the hardware limits. 


However, it is challenging to solve \eqref{eq:P1} directly. One reason is that it is a well-known NP hard problem to minimize the group sparsity ($\ell_0$ norm) in \eqref{eq:wgs}. Another reason is that the term $\sum_{f=1}^F \left(1-\frac{\tau_f}{\beta_{2,f}}\right)W_fR_{\ell}^f$ with $R_{\ell}^f$ \eqref{eq:mrtrate}  in a coupled structure with the transmit power is like the sum rate expression of the single-input and single-output (SISO) interference networks and also leads to a NP-hard problem in optimization. The goal of this work is to efficiently compute high-quality suboptimal solutions of Problem \eqref{eq:P1}.

\subsection{Problem Reformulation}
In order to make the problem \eqref{eq:P1} tractable, it is a common approach to relax a group sparsity $\ell_0$-norm to a mixed $\ell_2/\ell_1$ norm. The weighted group sparsity of a vector $\mat{x}$ in \eqref{eq:wgs} is approximately expressed as $||\mat{x}||_{0,\mat{w}}^{G, |\mat{x}_g|} \approx \sum_{g=1}^{G} w_g || \mat{x}_g ||_2$,
which is non-smooth but convex (its minimization is known as a group \emph{Lasso} problem). However, \cite{SparsitylogFan} and \cite{RecoveringSparsitylog} provided
a comparison of serval non-convex approximations of $\ell_0$ norm and suggested that the following $\log$-based approximation usually has a better sparse recovery performance
\begin{align}
||\mat{x}||_{0,\mat{w}}^{G,|\mat{x}_g|} 
= \lim_{\epsilon\rightarrow 0} \sum_{g=1}^G w_g \frac{\log(1+ \epsilon^{-1}\mat{1}^T\mat{x}_g)}{\log(1+\epsilon^{-1})} \approx \sum_{g=1}^G w_g\frac{\log(1+ \epsilon^{-1}\mat{1}^T\mat{x}_g)}{\log(1+\epsilon^{-1})}, \label{eq:l1l2norm2}
\end{align}
where $\epsilon$ in \eqref{eq:l1l2norm2} is set to be a very small constant. The simulations in the paper imply the choice of $\epsilon$ has a very slight influence on the performance.
 
Based on \eqref{eq:l1l2norm2} and \eqref{eq:backhaulrate}, the BS power consumption in \eqref{eq:totalpowermodel} approximately becomes
\begin{align}
\widehat{P}_{BS} 
=  &\sum_{k=1}^K P_k^{sleep_0} + \sum_{k=1}^K \sum_{f=1}^F P^f_{sp,k} \frac{\log(1+ \epsilon^{-1}\mat{t}_{k,f}^T\mat{p})}{\log(1+\epsilon^{-1})}  \nonumber \\
&+ \sum_{f=1}^F\left(1 - \frac{\tau_f}{\beta_{2,f}}\right)\sum_{k=1}^K{\frac{1}{\eta_k}\mat{t}_{k,f}^T \mat{p}} + P_{haul}\sum_{\ell=1}^L\frac{R_{\ell}(\mat{p})}{100~Mbit/s}, \label{eq:totalpowermodel3}
\end{align}
where  $\mat{t}_{k} \triangleq \mat{T}_{BS,k}^{T}\mat{1}, \mat{t}_{k,f} \triangleq \mat{T}_{BS,k}^{f,T}\mat{1}$ and $R_{\ell}(\mat{p})$ in \eqref{eq:mrtrate}. 

 

The average individual UE rate on FC $f$ in \eqref{eq:mrtrate} can be rewritten to a vector-form
\begin{align}
R_{\ell}^f = \log_2\left(1+\frac{\mat{\alpha}_{\mathcal{B}_{\ell},\ell}^{f,T} \mat{p}}{W_f\sigma^2 + \mat{\alpha}_{\mathcal{K},\overline{\ell}}^{f,T}\mat{p}}\right)  =\log_2\left(W_f\sigma^2 +{\mat{\alpha}_{\mathcal{K},\ell}^{f,T} \mat{p}} \right) - \log_2\left({W_f\sigma^2 + \mat{\alpha}_{\mathcal{K},\overline{\ell}}^{f,T}\mat{p}}\right), \label{eq:MRTratenewb}
\end{align}
where $\mat{\alpha}_{\mathcal{B}_{\ell},\ell}^{f}$ is a $LF|\mathcal{B}_{\ell}| \times 1$ all-zeros vector except for the corresponding positions of $\{ \alpha_{k,\ell}^f (\delta_{k,\ell}^f(N_k-1)+1)\}_{k\in\mathcal{B}_{\ell}}$, and $\mat{\alpha}_{\mathcal{K},\overline{\ell}}^{f}$ is similarly defined. In \eqref{eq:MRTratenewb}, we define $\mat{\alpha}_{\mathcal{K},\ell}^{f} \triangleq \mat{\alpha}_{\mathcal{B}_{\ell},\ell}^{f} +  \mat{\alpha}_{\mathcal{K},\overline{\ell}}^{f}$.
Observe that ${R_{\ell}^f}$ in \eqref{eq:MRTratenewb} is a difference of two concave (DC) functions of $\mat{p}$. 


Based on the reformulation in \eqref{eq:totalpowermodel3} and in \eqref{eq:MRTratenewb}  of the rate constraint and objective function, respectively, after moving the constant terms in the objective function Problem \eqref{eq:P1} becomes 
\begin{subequations} \label{eq:P21}
\begin{align}
\min_{\mat{p}\geq \mat{0}}~& \sum_{k=1}^K \sum_{f=1}^F P^f_{sp,k} {\log(\epsilon + \mat{t}_{k,f}^T\mat{p})} + \sum_{f=1}^F\left(1 - \frac{\tau_f}{\beta_{2,f}}\right)\sum_{k=1}^K{\frac{1}{\eta_k}\mat{t}_{k,f}^T \mat{p}} \label{eq:P2a1}  \\
\mathrm{s.t.}~
&\sum_{f=1}^F\left(1 - \frac{\tau_f}{\beta_{2,f}}\right)W_f\left( \log_2\left(W_f\sigma^2 +{\mat{\alpha}_{\mathcal{K},\ell}^{f,T} \mat{p}} \right) - \log_2\left({W_f\sigma^2 + \mat{\alpha}_{\mathcal{K},\overline{\ell}}^{f,T}\mat{p}}\right)\right) \geq {\gamma}_{\ell},~\forall \ell \in \mathcal{L}   \label{eq:P2b1} \\
&\mat{t}_{k}^T\mat{p} \leq {P}_{BS,k}^{max},~\forall k \in \mathcal{K},   \label{eq:P2c1}
\end{align} 
\end{subequations}
where the total backhaul power consumption term is removed in \eqref{eq:P2a1}, because the rate constraint \eqref{eq:P2b1} will be optimally achieved with "equality", i.e., $R_{\ell}(\mat{p})=\gamma_{\ell}$ (constant term). 
However, Problem \eqref{eq:P21} is still difficult to solve, since it is a \emph{concave-minimization} problem with the \emph{DC constraints}.

\section{SCA-based Algorithms and Solutions} \label{sec:algorithm}

In this section, the SCA-based algorithm is applied to compute the locally optimal solutions of the non-convex problem \eqref{eq:P21}.
The basic idea of the SCA-based algorithm (in spirit of \cite{SCA2,SCA4}) is to iteratively 1) construct a surrogate function as the upper bound of the objective/constraint function at the current solution and then 2) optimize the problem with surrogate functions which yields the next estimation of the parameters. 


\subsection{Technical Preliminaries}
Consider the following non-convex optimization problem:
\begin{subequations}  \label{eq:P2}
\begin{align}
\min_{\mat{x} \in \mathbb{R}^M} ~&y(\mat{x}) \label{eq:P2a}  \\
\mathrm{s.t.}~~
&c_j(\mat{x}) \leq 0,~  j =1,\cdots,J,~~ \mat{x} \in \Omega,   \label{eq:P2b}
\end{align} 
\end{subequations}
where $y, c_j: \mathbb{R}^M \rightarrow \mathbb{R}$ are non-convex but smooth functions with the form of 
\begin{align}
y(\mat{x}) \triangleq y^+(\mat{x}) - y^-(\mat{x})~\mathrm{and}~c_j(\mat{x}) \triangleq c_j^+(\mat{x}) - c_j^-(\mat{x}),~~\forall j
\end{align}
where $y^+, y^-, c_j^+, c_j^-: \mathbb{R}^M \rightarrow \mathbb{R}$ are continuous convex functions, and $\Omega$ is a convex set in $\mathbb{R}^M$. We define $\mathcal{X} \triangleq \{\mat{x}\in \Omega: c_j(\mat{x})\leq 0,~ j=1,\cdots, J\}$.

Problem \eqref{eq:P2} is a \emph{DC program with DC constraints} (non-convex in general). 
By the SCA, a common scheme to generate a surrogate function is to \emph{linearize} the non-convex functions by using a first-order Taylor series. For example, either the \emph{completely linearized (CL)} function
\begin{align}
y^{CL}(\mat{x}, \mat{z}) = y(\mat{z}) + (\nabla y(\mat{z}))^T(\mat{x}-\mat{z})  \label{eq:lineara}
\end{align}
or the \emph{partially linearized (PL)} function
\begin{align}
y^{PL}(\mat{x}, \mat{z}) = y^+(\mat{x}) -  \left(y^-(\mat{z}) + (\nabla y^-(\mat{z}))^T(\mat{x}-\mat{z})\right)  \label{eq:linearb}
\end{align}
can be the surrogate function of $\mat{y}(\mat{x})$, which is tight at a feasible point $\mat{z}$, i.e.,
  \begin{subnumcases}
    {y^{CL}(\mat{x}, \mat{z})~\mathrm{and}~y^{PL}(\mat{x}, \mat{z})}
= y(\mat{x}) & $\mathrm{when}~\mat{x}=\mat{z}$   \label{eq:lineaConeq}\\  
\geq  y(\mat{x})  & $\mathrm{otherwise}$. \label{eq:lineaCongeq}
\end{subnumcases}


Similarly, $c_j^{CL}(\mat{x})$ or $c_j^{PL}(\mat{x})$ is assumed to be the surrogate function of the DC constraint function $c_j(\mat{x}), \forall j$.
Then, the DC program with DC constraints can be approximately formulated as a sequence of convex optimization problems (in multiple iterations), and each can be solved using algorithms and toolbox from convex optimization theory. Therefore, Problem \eqref{eq:P2} can be suboptimally but efficiently solved by the following 
Algorithm \ref{al:DC1} and its variants. 

\begin{algorithm}
Initialization: $i = 0$, $\mat{x}^{(0)}\in \mathcal{X}$ and $\epsilon_{th}$.

\Repeat{$||\mat{x}^{(i)}- \mat{x}^{(i-1)}|| \leq \epsilon_{th}$}
{

Generate the surrogate functions $\mat{y}^{PL}(\mat{x}, \mat{x}^{(i)})$ and $\mat{c}_j^{PL}(\mat{x}, \mat{x}^{(i)})$ by following \eqref{eq:lineara};

Solve the convex optimization problem
\begin{align}
\mat{x}^{(i+1)} = \arg \min_{\substack{x\in \Omega, \\ c_j^{PL}(\mat{x}, \mat{x}^{(i)})\leq 0, ~j=1,\cdots,J}} y^{PL}(\mat{x}, \mat{x}^{(i)}); \label{eq:linearAl1}
\end{align}

$ i \leftarrow i+1$. 

 }
 
\caption{SCA-based Algorithm to Solve DC Program \eqref{eq:P2}}\label{al:DC1}
\end{algorithm}

\begin{remark}
In principle, both PL functions and the CL functions (if they are feasible) can be flexibly used as the surrogate functions of the non-convex objective and constraint functions, which might lead to some variants of Algorithm \ref{al:DC1}. 
\end{remark}



\subsection{Solutions of BS Energy Consumption Minimization}
By the above SCA-based algorithm, Problem  \eqref{eq:P21} as a DC program can be solved.

At a feasible point $\mat{q}$, based on \eqref{eq:lineara} and \eqref{eq:linearb} and after removing the constant terms, the surrogate function of the concave objective function and the DC constraint in  \eqref{eq:P2} can be expressed by
\begin{align}
&\widehat{P}_{BS}^{S}(\mat{p},\mat{q}) \triangleq
\sum_{k=1}^K \sum_{f=1}^F P^f_{sp,k} \frac{\mat{t}_{k,f}^T\mat{p}}{\epsilon + \mat{t}_{k,f}^T\mat{q}} + \sum_{f=1}^F\left(1 - \frac{\tau_f}{\beta_{2,f}}\right)\sum_{k=1}^K{\frac{1}{\eta_k}\mat{t}_{k,f}^T \mat{p}}, \label{eq:P2obja}  \\
&R_{\ell}^{S}(\mat{p}, \mat{q}) \triangleq \sum_{f=1}^F(1-\frac{\tau_f}{\beta_{2,f}})W_f  \Big(\log_2\left(\frac{W_f\sigma^2 +{\mat{\alpha}_{\mathcal{K},\ell}^{f,T} \mat{p}}}{W_f\sigma^2 +{\mat{\alpha}_{\mathcal{K},\overline{\ell}}^{f,T} \mat{q}} } \right) - \frac{1}{\log(2)}\cdot\frac{\mat{\alpha}_{\mathcal{K},\overline{\ell}}^{f,T}(\mat{p}-\mat{q})}{W_f\sigma^2 + \mat{\alpha}_{\mathcal{K},\overline{\ell}}^{f,T}\mat{q}}\Big) \geq \gamma_{\ell}      \label{eq:P2objb},
\end{align} 
respectively. 

After selecting a feasible initial point $\mat{p}^{(0)}$, Problem \eqref{eq:P21} can be suboptimally solved by the following Algorithm \ref{al:DCPower1}.

\begin{algorithm}
Initialization: $i = 0$, a feasible $\mat{p}^{(0)}$ and $\epsilon_{th}$.

\Repeat{$||\mat{p}^{(i)}- \mat{p}^{(i-1)}||_2 \leq \epsilon_{th}$}
{


Solve the convex optimization problem
\begin{align}
\mat{p}^{(i+1)} = \arg \min_{\substack{\mat{p}\geq \mat{0}, ~\mat{t}_{BS,k}^T\mat{p}\leq P_{BS,k}^{max},~\forall k\in \mathcal{K} \\ R_{\ell}^S(\mat{p}, \mat{p}^{(i)})\geq \gamma_{\ell}, ~\forall \ell \in \mathcal{L}}} \widehat{P}_{BS}^S(\mat{p}, \mat{p}^{(i)}); \label{eq:DC2a}
\end{align}

$ i \leftarrow i+1$. 

 }
 
\caption{SCA-based Algorithm to Solve Problem \eqref{eq:P21}}\label{al:DCPower1}
\end{algorithm}
In Algorithm \ref{al:DCPower1}, \eqref{eq:DC2a} is a convex optimization problem with a linear objective function and convex constraints, 
which can be optimally and efficiently solved by the CVX toolbox.  

\begin{remark}
The surrogate function $R_{\ell}^S(\mat{p}, \mat{p}^{(i)})$ in \eqref{eq:P2objb} is an upper bound of the real rate function $R_{\ell}(\mat{p})$, but in each iteration it is always achieved that  $R_{\ell}^S(\mat{p}^\star, \mat{p}^{(i)}) = \gamma_{\ell}, \forall \ell$ where $\mat{p}^{\star}$ is the optimal solution to \eqref{eq:DC2a} because of $R_{\ell}^S(\mat{p}^\star, \mat{p}^{(i)}) = R_{\ell}(\mat{p}^\star, \mat{p}^{(i)})  = \gamma_{\ell}, \forall \ell$ (implied by \eqref{eq:lineaConeq}). This makes that each UE rate requirement can be finally guaranteed.  
\end{remark}

\begin{proposition} \label{pr:KKT}
The SCA-based algorithm in Algorithm \ref{al:DCPower1} always converges to a KKT stationary solution of Problem \eqref{eq:P21}. 
\end{proposition}
\begin{IEEEproof}
See Appendix \ref{subsec:ProofConv}. 
\end{IEEEproof}

Therefore, a local-optimal solution $\overline{\mat{p}}$ to Problem \eqref{eq:P21} can be obtained by Algorithm \ref{al:DCPower1}, which is not guaranteed to be global optimal. Then, this solution also gives the answers to the problems \emph{P1-P4} in Section \ref{sec:introcontribution}.

\subsection{Performance Analysis}\label{sec:performance}

We compare our proposed algorithm based on the flexible assumptions \emph{A2-A4} in Section \ref{sec:introcontribution} with some baselines that study the same BSs power minimization problem with the proposed BS power model but based on the assumptions \emph{R2-R5} in Section \ref{sec:relatedwork} in a theoretical way. 
\begin{proposition}
Based on the flexible system assumptions {A2-A4} in Section \ref{sec:introcontribution}, our proposed green resource management mechanism always outperforms those baselines which are based on the assumptions {R2-R5} in Section \ref{sec:relatedwork}.
\end{proposition}
\begin{IEEEproof}
Similar to Definition \ref{de:BSpower}, we let $\mat{p}_{UE,\ell}  \in \mathbb{R}^{F|\mathcal{B}_{\ell}|\times 1}$, $\mat{p}_{UE,\ell}^f \in \mathbb{R}^{|\mathcal{B}_{\ell}|\times 1}$, and $\mat{p}_{FC,f} \in \mathbb{R}^{L|\mathcal{B}_{\ell}|\times 1}$ denote the power of the BSs set $\mathcal{B}_{\ell}$ to UE $\ell$ on all FCs, the power of the BSs set $\mathcal{B}_{\ell}$ to UE $\ell$ on FC $f$, and the power of all the BSs to all the UEs on FC $f$, respectively.  The "restricted" assumptions \emph{R2-R5} can be equivalently formulated to the following theoretical constraints
\begin{align}
&\mathrm{Assumption}~\emph{R2}~\Leftrightarrow~ ||\mat{p}_{BS,k}^f||_0 \leq 1,~~\forall k\in \mathcal{K}, \forall f\in\mathcal{F},  \label{eq:as1} \\
&\mathrm{Assumption}~\emph{R3}~\Leftrightarrow~ ||\mat{p}_{UE,\ell}||_0^{|\mathcal{B}_{\ell}|, F} = 1, ~~\forall \ell \in \mathcal{L},   \label{eq:as2} \\
&\mathrm{Assumption}~\emph{R4}~\Leftrightarrow~ ||\mat{p}_{UE,\ell}||_0^{F, |\mathcal{B}_{\ell}|} = 1, ~~\forall k \in \mathcal{K},    \label{eq:as3}  \\
&\mathrm{Assumption}~\emph{R5}~\Leftrightarrow~ ||\mat{p}_{FC,f}||_0^{L,|\mathcal{B}_{\ell}|} \leq 1,~~\forall f\in \mathcal{F},   \label{eq:as4}
\end{align}
respectively. Therefore, for example, one baseline with assumption \emph{R2} can be formulated to the optimization problem \eqref{eq:P1} but with an extra constraint \eqref{eq:as1}. In optimization, more constraints used for the same objective optimization problem will degrade the performance (or have the same performance when this extra constraint is inactive), since the feasible solution set is shrunk.
In this work, these constraints \eqref{eq:as1}-\eqref{eq:as4} have been, in fact, relaxed by the general assumptions \emph{A2-A4} as shown in Problem \eqref{eq:P1} , and thus its outperformance is verified. 
\end{IEEEproof}

\subsection{Implementation}

The implementation of the proposed semi-dynamic green resource management mechanism during each A-LSF in a HetNet is summarized as follows.
\begin{itemize}
\item \textbf{Step 1 ({LSF Acquisition})}: At the beginning of an A-LSF, the CP collects the predicted LSF values of the network;
\item \textbf{Step 2 ({Green Resource Management})}: Based on the LSF values, the CP solves Problem \eqref{eq:P21} by Algorithm \ref{al:DCPower1}. According to the group sparse vector $\overline{\mat{p}}$ that is obtained, the BS-UE association, BS/UE-FC assignment, downlink transmit power allocation and the sleep modes for BSs can be jointly determined;
\item \textbf{Step 3a ({CSI Estimation})}: At the beginning of each coherence time, each UE transmits the uplink training sequences to its associated BSs on the assigned FCs, based on which each BS estimates its local CSI of its serving UEs; 
\item \textbf{Step 3b ({MRT Beamforming Design})}: Each BS locally designs the MRT beamforming vectors for its serving UEs on the assigned FC based on the estimated CSI in \textbf{Step 3a} and the transmit power vector $\overline{\mat{p}}$ in \textbf{Step 2};
\item \textbf{Step 3c ({Downlink Transmission})}: Each BS transmits the desired data symbols to its serving UEs by the same MRT beamforming vectors until the end of the coherence time;
\item \textbf{Step 4}:  Repeat \textbf{Step 3a} to \textbf{Step 3c} until the end of the A-LSF.
\end{itemize}


\section{Simulations and Discussions} \label{sec:simulation}

In this section, the performance of the proposed algorithm is evaluated on a 3-macro cell two-tier HetNet. Each macro cell is a regular hexagon with a radius of 250 meters and a single macro BS located at the center, where the same number of pico BSs and UEs are randomly deployed within each macro cell with the simulation parameters in Table \ref{table:systempara}. 

As shown in Section \ref{sec:performance}, we have already proved that our proposed algorithm always outperforms the baselines based on the restricted BS-UE association and BS/UE-FC assignment assumption \emph{R2-R5} in Section \ref{sec:relatedwork}, and thus the focus herein is on two other  baselines:
\begin{itemize}
\item $L_{2,1}$ Approx: It denotes the performance of the same optimization by Algorithm \ref{al:DCPower1} but using the $\ell_{1}/\ell_2$ mixed norm to approximate the $\ell_0$ norm instead of \eqref{eq:l1l2norm2}. This baseline is to show the impact of the $\ell_0$ norm approximation;  
\item Min. T-Power: This baseline is generated when only the downlink transmit power is minimized, which is a quite widely-used metric in the previous work on energy efficiency/saving. 
\end{itemize}
\begin{table}\caption{HetNet system parameters}
\centering
\begin{tabular}{ c | l || c |l} 
\hline
Pilot length & Total No. of UEs & Spectrum Bands  & 783-803 MHz, 1900-1920 MHz  \\
\hline	
$P_{sp,k}^f$  & Reference \cite{ParameterBSPower} & $P_{k}^{sleep_0}$ & 75 Watt (macro), 4.3 Watt (pico)  \\
\hline 
 $\sigma^2$ & -174 dBm/Hz & $P_{BS,k}^{max}$ & 40 Watt (macro), 1 Watt (pico)  \\ 
\hline
 Path Loss  & Reference \cite{CarrieronPathloss} & PA efficiency & 35\% (macro), 25\% (pico) \\ 
 \hline
\end{tabular}
\label{table:systempara}
\end{table}

\subsection{Performance Evaluation for Deterministic UEs}
We first evaluate the performance of Algorithm \ref{al:DCPower1} within a typical A-LSF, where the UEs' locations can be considered to be fixed because the LSF is not varying during each LSF time period. We assume $5$ pico BSs per macro cell. The partial loaded scenario is considered, where $5$ UEs are located within each macro cell and each UE has a 2 Mbits/s data rate requirement. No pilot sequence is reused. As shown in Table \ref{table:systempara}, a total 40 MHz spectrum is available. 

\begin{figure}[t]
         \centering
           \includegraphics[width=1.0\textwidth]{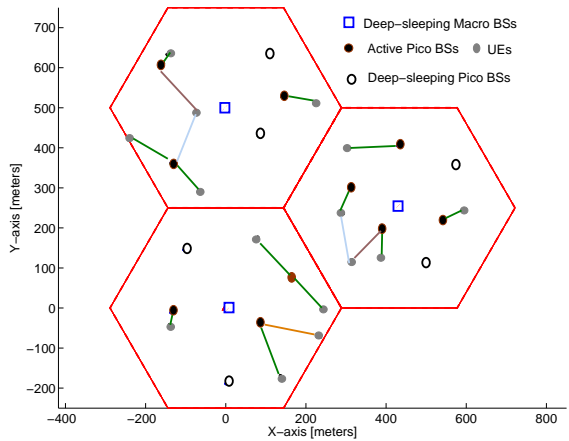}
                    \caption{A simulation result for Algorithm \ref{al:DCPower1} with a per UE rate requirement 2 Mbits/s.}\label{fig:example}
   \end{figure}

A result example for Algorithm \ref{al:DCPower1} is shown in Fig. \ref{fig:example}, where each two 20 MHz spectrum band is equally spit into 4 FCs
for the spectrum band 783-803 MHz, 1900-1920 MHz, respectively. We observe that all macro BSs are in deep-sleep mode as well as some pico BSs because of the low load. Another interesting observation is that most UEs prefer to reuse the FCs with low frequency (in green color) which are with less path loss. The BS-UE association, BS/UE-FC assignment and BSs deep-sleep status are clearly illustrated.

\begin{figure}[t]
         \centering
           \includegraphics[width=.6\textwidth]{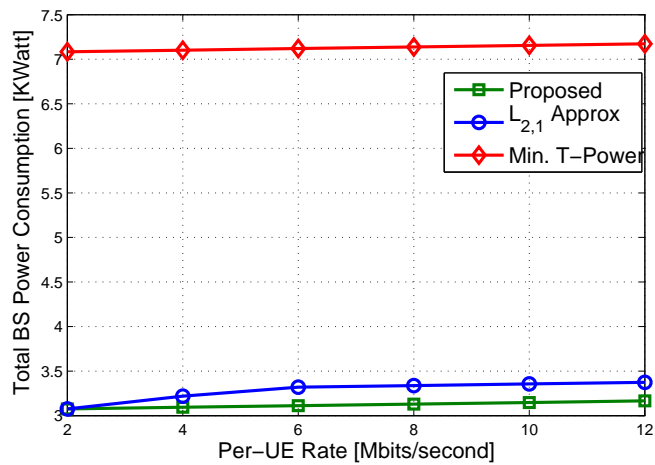}
                    \caption{Total BS power consumption v.s. UE rate requirement}\label{fig:powerconsumption1}
   \end{figure}

When we assume two FCs are adopted where each FC has a bandwidth of 20 MHz, .e.g, the multiple access scenario, an energy consumption comparison with the baselines "$L_{2,1}$ Approx" and "Min. T-Power" is shown in Fig. \ref{fig:powerconsumption1}. Observe that the energy consumption is increasing with the UE's rate requirement and our algorithm can achieve a more than 50\% energy reduction compared with  the "Min. T-Power", since the "Min. T-Power" does not optimize the sleep modes. This implies that our proposed flexible BS power model provides more freedoms for further energy saving. The $\log$-based approximation also outperforms the $\ell_{1}/\ell_2$ mixed norm. 

\begin{figure}[t]
         \centering
           \includegraphics[width=.6\textwidth]{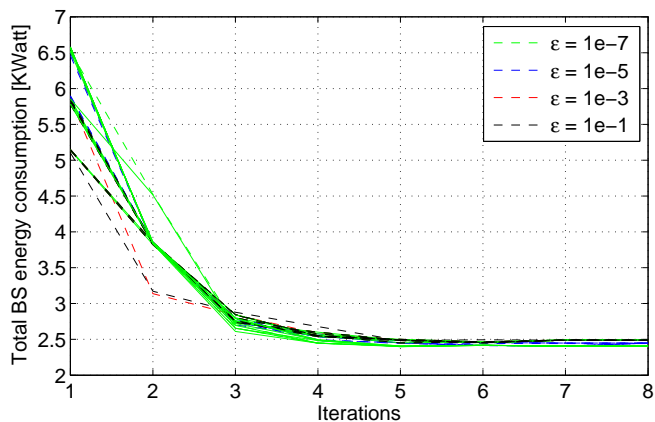}
                    \caption{Convergence performance of Algorithm \ref{al:DCPower1} }\label{fig:covergece}
   \end{figure}

In Fig. \ref{fig:covergece} the convergence behavior of Algorithm \ref{al:DCPower1} is shown, where we set the parameter  $\epsilon$ for the $\ell_0$ norm approximation in \eqref{eq:l1l2norm2} as $\epsilon \in \{10^{-1}, 10^{-3}, 10^{-5}, 10^{-7}\}$, for each $\epsilon$ 10 random initializations are used. It is shown in Fig. \ref{fig:covergece} that the used $\ell_0$ norm approximation in \eqref{eq:l1l2norm2} is robust to the choice of $\epsilon$ and different initializations might lead to different KKT stationary solutions with similar convergence rate. 


\subsection{Average Performance Evaluation}
The average performance of the proposed algorithm is evaluated by 100 Monte Carlo simulations, where the locations of $5$ UEs are randomly generated within each macro cell.

\begin{figure}[t]
         \centering
           \includegraphics[width=.6\textwidth]{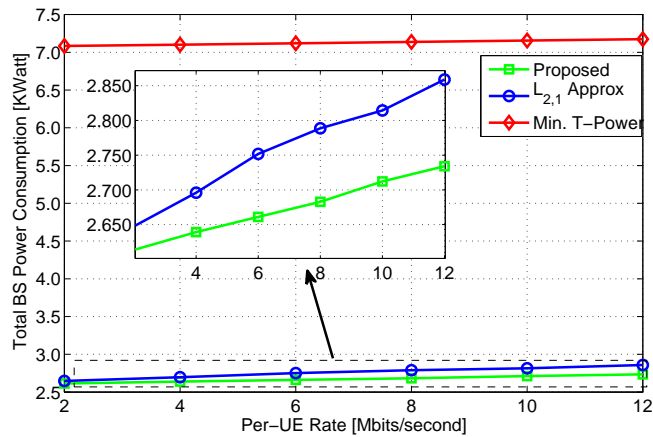}
                    \caption{Average total BS power consumption performance vs.  UE rate}\label{fig:averageEEs}
   \end{figure}

The average energy consumption for 2 FCs with respect to the UEs' data rate requirement is shown in Fig. \ref{fig:averageEEs}, which has a similar behavior (also with more than $50\%$ energy reduction) with deterministic scenario in Fig. \ref{fig:powerconsumption1}. This implies that the performance of the proposed algorithm is not highly influenced by the specific channel values. 

\begin{figure}[t]
         \centering
           \includegraphics[width=.6\textwidth]{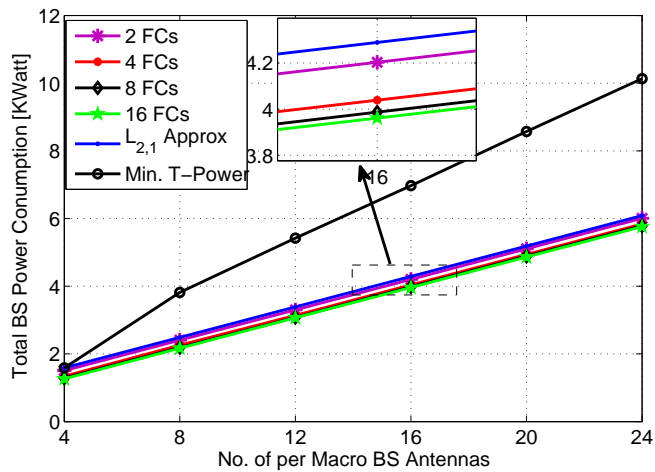}
                    \caption{Average total BS power consumption vs. macro BS antennas and FC splitting}\label{fig:FCs}
   \end{figure}

In Fig. \ref{fig:FCs}, we illustrate the total energy consumption for the higher UE rate requirement of 20 Mbits/s, where some of macro BSs are not in deep-sleep, and thus the signal processing power scales with the number of macro BS antennas. Note that the signal processing power also depends on the bandwidth of the assigned FCs. By the sparse power control, the proposed algorithm and the "$L_{2,1}$ Approx" can slow down this increase by reducing the assigned bandwidth compared with the "Min. T-Power".
With this respect, the impact of the bandwidth of each individual FC on the energy consumption is evaluated, where each 20 MHz spectrum band is equally split into $2$, $4$ and $8$ FCs, respectively. The results show that the energy consumption by the proposed Algorithm \ref{al:DCPower1} is slightly decreasing with the carrier splitting, while the two baselines seem to not be sensitive to the amount carrier splitting. In contrast to carrier aggregation, narrowing a FC will sacrifice the spectrum efficiency, but a more flexible resource usage for scheduling is allowed. This is in particular important for the partial-loaded scenarios, where the wide carrier may be not necessary. The study of the trade-off between the spectrum efficiency and the energy efficiency with respect to the bandwidth and the number of FCs will be done in our future work.



\section{Conclusions}

In this paper, motivated by the requirement for energy saving in the partially loaded HetNets, we propose an optimization scheme for the system operation to be as flexible and scalable as possible. This flexibility provides more freedom to help the network reduce the energy consumption by deactivating some unnecessary hardware components.
A flexible BS power consumption model is developed to support the scalability, 
which allows the BS to control the system resources, such as antennas and frequency carriers, for energy saving by group sparse power control. Based on this power model, a BS energy consumption minimization problem while supporting each user's rate requirement is formulated and optimized only with respect to a transmit power vector. 
Solving this problem yields solutions for a series of green resource management problems, such as BS-UE association, BS/UE-FC assignment, the BS signal processing power levels, and the energy minimization can be jointly solved. In addition, this work provides a general framework for BS energy minimization, which is independent of the BSs tiers/density and the number/bandwidth of the FCs. 
Simulation results indicate that the proposed algorithm is capable of reducing the BS power consumption by more than $50\%$.

\appendices

\section{Proof of Lemma \ref{le:averagerate}} \label{subsec:ProofRate}

\begin{IEEEproof}
With the MRT beamforming $\mat{w}_{k,\ell}^f = \sqrt{p_{k,\ell}^f}\overrightarrow{\hat{\mat{h}}}_{k,\ell}^f, \forall k, \ell, f$, \eqref{eq:rate1} becomes
\begin{align} 
R_{\ell}^{f} &= \mathbb{E}_{\tilde{\mat{h}}}\Bigg\{\log_2\Bigg(1+\frac{\sum_{k \in \mathcal{B}_{\ell}} p_{k,\ell}^f \alpha_{k,\ell}^f |\tilde{\mat{h}}_{k, \ell}^{f,H}\overrightarrow{\hat{\mat{h}}}_{k, \ell}^f |^2 }{ \substack{W_f\sigma^2 + {\sum_{\overline{k}\in \mathcal{K}\backslash\mathcal{B}_{\ell}} \sum_{j \in \mathcal{U}_{\overline{k}}} p_{\overline{k},j}^f\alpha_{\overline{k},\ell}^f |\tilde{\mat{h}}_{\overline{k}, \ell}^{f,H} \overrightarrow{\hat{\mat{h}}}_{\overline{k},j}^f|^2} \\ + {\sum_{k\in\mathcal{B}_{\ell}}\sum_{\overline{\ell} \in \mathcal{U}_{k}\backslash\{\ell\}} p_{k,\overline{\ell}}^f \alpha_{k,{\ell}}^f |\tilde{\mat{h}}_{k, \ell}^{f,H}(t) \hat{\mat{h}}_{k,\overline{\ell}}^f(t)|^2}}}\Bigg)\Bigg\} \label{eq:rate2} \\
&\approx \log_2\Bigg(1+\frac{\sum_{k \in \mathcal{B}_{\ell}} p_{k,\ell}^f \alpha_{k,\ell}^f \mathbb{E}_{\tilde{\mat{h}}}\{|\tilde{\mat{h}}_{k, \ell}^{f,H}\overrightarrow{\hat{\mat{h}}}_{k, \ell}^f |^2\} }{\substack{W_f\sigma^2 + {\sum_{\overline{k}\in \mathcal{K}\backslash\mathcal{B}_{\ell}} \sum_{j \in \mathcal{U}_{\overline{k}}} p_{\overline{k},j}^f\alpha_{\overline{k},\ell}^f \mathbb{E}_{\tilde{\mat{h}}}\{|\tilde{\mat{h}}_{\overline{k}, \ell}^{f,H} \overrightarrow{\hat{\mat{h}}}_{\overline{k},j}^f|^2\}} \\ + {\sum_{k \in \mathcal{B}_{\ell}}\sum_{\overline{\ell} \in \mathcal{U}_{k}\backslash\{\ell\}} p_{k,\overline{\ell}}^f \alpha_{k,{\ell}}^f \mathbb{E}_{\tilde{\mat{h}}}\{|\tilde{\mat{h}}_{k, \ell}^{f,H}(t) \hat{\mat{h}}_{k,\overline{\ell}}^f(t)|^2\}}}}\Bigg) \label{eq:rate2a} \\
&= \log_2\Bigg(1+\frac{\sum_{k \in \mathcal{B}_{\ell}} p_{k,\ell}^f(1-\delta_{k,\ell}^f)\alpha_{k,\ell}^fN_{k} }{ W_f\sigma^2 + {\sum_{\overline{k}\in \mathcal{K}\backslash\mathcal{B}_{\ell}} \sum_{j \in \mathcal{U}_{\overline{k}}} p_{\overline{k},j}^f \alpha_{\overline{k}, \ell}^f} + {\sum_{k \in \mathcal{B}_{\ell}}\sum_{\overline{\ell} \in \mathcal{U}_{k}\backslash\{\ell\}} p_{k,\overline{\ell}}^f \alpha_{k,\ell}^f}}\Bigg)
\label{eq:rate3}
 \end{align} 
where 
\eqref{eq:rate2a} is derived based on the approximation $\mathbb{E}_x\{\log_2(1+\frac{f_1(x)}{f_2(x)})\} \approx \log_2(1+\frac{ \mathbb{E}_x\{f_1(x)\}}{ \mathbb{E}_x\{f_2(x)\}})$, and \eqref{eq:rate3} is derived based on the following results:
\begin{align}
\mathbb{E}_{\tilde{\mat{h}}}\{|\tilde{\mat{h}}_{k,\ell}^f \overrightarrow{\hat{\mat{h}}}_{k,\ell}^f|^2\} &= \mathbb{E}\{|\left(\hat{\mat{h}}_{k,\ell}^f + \mat{e}_{k,\ell}^f\right)^H \overrightarrow{\hat{\mat{h}}}_{k,\ell}^f|^2\}      \label{eq:s1}  \\
&= \mathbb{E}\{|\hat{\mat{h}}_{k,\ell}^{f,H} \overrightarrow{\hat{\mat{h}}}_{k,\ell}^f|^2\} +  \mathbb{E}\{{\mat{e}}_{k,\ell}^{f,H} \overrightarrow{\hat{\mat{h}}}_{k,\ell}^f|^2\} \label{eq:s2} \\
&= \mathbb{E}\{||\hat{\mat{h}}_{k,\ell}^{f}||^2\}  +  \overrightarrow{\hat{\mat{h}}}_{k,\ell}^{f,H}\mathbb{E}\{|{\mat{e}}_{k,\ell}^{f}{\mat{e}}_{k,\ell}^{f,H}\}\overrightarrow{\hat{\mat{h}}}_{k,\ell}^f     \label{eq:s3} \\
&=  \delta_{k,\ell}^f N_k  +  (1-\delta_{k,\ell}^f), ~~\forall k \in \mathcal{B}_{\ell} \label{eq:s4}
\end{align}
where \eqref{eq:s1} is based on the estimated channel model in \eqref{eq:channelesti}, and \eqref{eq:s2} is based on the fact $\mathbb{E}\{ \hat{\mat{h}}_{k,\ell}^{f,H} \overrightarrow{\hat{\mat{h}}}_{k,\ell}^f\overrightarrow{\hat{\mat{h}}}_{k,\ell}^{f,H}{\mat{e}}_{k,\ell}^{f}\}=0$ because  ${\mat{e}}_{k,\ell}^{f}$ is zero-mean Gaussian and is independent of $\hat{\mat{h}}_{k,\ell}^{f}$, and \eqref{eq:s4} is based on the derived result in \eqref{eq:channelesti}.

The average inter-BS interference terms in the denominator of \eqref{eq:rate2a} are derived to
\begin{align}
\mathbb{E}_{\tilde{\mat{h}}}\left\{ |\tilde{\mat{h}}_{\overline{k}, \ell}^{f,H} \overrightarrow{\hat{\mat{h}}}_{\overline{k},j}^f|^2\right\}  
&= \overrightarrow{\hat{\mat{h}}}^{f,H}_{\overline{k},j} \mathbb{E}_{\tilde{\mat{h}}} \{ \tilde{\mat{h}}_{\overline{k}, \ell}^f \tilde{\mat{h}}_{\overline{k}, \ell}^{f,H}\} \overrightarrow{\hat{\mat{h}}}^f_{\overline{k},j} =1,~~\forall j \in \mathcal{U}_{\overline{k}},~\overline{k} \in \mathcal{K}\backslash\{\mathcal{B}_{\ell}\} \label{eq:Interf}  \\
\mathbb{E}_{\tilde{\mat{h}}}\left\{|\tilde{\mat{h}}_{k, \ell}^{m,H} \overrightarrow{\hat{\mat{h}}}_{k,\overline{\ell}}^m|^2\right\} &=  1,~~\forall {\overline{\ell} \in \mathcal{U}_{k}\backslash\{\ell\}},~\overline{k} \in \mathcal{B}_{\ell}
\end{align}
where \eqref{eq:Interf} is because $\tilde{\mat{h}}_{\overline{k},\ell}^f$ is unit-variance Gaussian and is independent of $\overrightarrow{\hat{\mat{h}}}_{\overline{k},\overline{\ell}}^f, \forall \overline{\ell} \neq \ell$.
\end{IEEEproof}

\section{Proof of Proposition \ref{pr:KKT}}\label{subsec:ProofConv}
\begin{IEEEproof}
The proof of Proposition \ref{pr:KKT} has two aspects: 1) the convergence of Algorithm \ref{al:DCPower1} and 2)  the solutions of Algorithm \ref{al:DCPower1}. 

\emph{1) Convergence:} The convergence of the algorithm is implied by the fact that the iterative sequence $\{\widehat{P}_{BS}(\mat{p}^{(i)})\}_{i=1}^{+\infty}$ is monotonically decreasing. At the $i$-th iteration, we have
\begin{align}
\widehat{P}_{BS}(\mat{p}^{(i+1)}) \stackrel{(a)}{=} \widehat{P}_{BS}(\mat{p}^{(i+1)}, \mat{p}^{(i+1)}) \stackrel{(b)}{\leq}  \widehat{P}_{BS}(\mat{p}^{(i+1)}, \mat{p}^{(i)}) \stackrel{(c)}{\leq} \widehat{P}_{BS}(\mat{p}^{(i)}, \mat{p}^{(i)}) \stackrel{(d)}{=} \widehat{P}_{BS}(\mat{p}^{(i)}),
\end{align}
where both the equalities $\mathrm{(a)}$ and $\mathrm{(d)}$ are based on \eqref{eq:lineaConeq}, and the inequalities $\mathrm{(b)}$ and $\mathrm{(c)}$ are based on \eqref{eq:lineaCongeq} and the convex optimization of \eqref{eq:DC2a} (optimal updating).  Considering that the constraints form a closed set, there exists a cluster point of the sequence $\{\widehat{P}_{BS}(\mat{p}^{(i)})\}_{i=1}^{+\infty}$. Let $\overline{\mat{p}} \triangleq \lim_{i\rightarrow +\infty} \mat{p}^{(i)}$ be the cluster point solution returned by Algorithm \ref{al:DCPower1} with a sufficiently small $\epsilon_{th}$. 

\emph{2) KKT Solutions:} We will show the cluster point solution $\overline{\mat{p}}$ is a KKT stationary point of the original problem \eqref{eq:P21}. Considering the properties of the cluster point, we have $\mat{p}^{(i)}=\mat{p}^{(i+1)} = \overline{\mat{p}}$ with $i\rightarrow +\infty$ for the optimization of \eqref{eq:DC2a}. Therefore, given $\mat{p}^{(i)}= \overline{\mat{p}}$, the optimal solution $\mat{p}^{(i+1)} =\overline{\mat{p}}$ of \eqref{eq:DC2a} should satisfy the following KKT conditions
 \begin{subequations}
\begin{align}
 &\sum_{k=1}^K \sum_{f=1}^F P^f_{sp,k} \frac{\mat{t}_{k,f}^T}{\epsilon + \mat{t}_{k,f}^T\overline{\mat{p}}} + \sum_{k=1}^K\theta_k\mat{t}_{BS,k}^T \nonumber \\
&~~~~~~~~~~+\sum_{\ell=1}^L\zeta_{\ell}\sum_{f=1}^F\left(1 - \frac{\tau_f}{\beta_{2,f}}\right)\frac{W_f}{\log(2)}\left(\frac{\mat{\alpha}_{\mathcal{K},\ell}^{f,T}}{W_f\sigma^2 +{\mat{\alpha}_{\mathcal{K},\ell}^{f,T} \overline{\mat{p}}}}  - \frac{\mat{\alpha}_{\mathcal{K},\overline{\ell}}^{f,T}}{W_f\sigma^2 + \mat{\alpha}_{\mathcal{K},\overline{\ell}}^{f,T}\overline{\mat{p}}}\right)
  = 0 \label{eq:KKTa}\\
&0\leq \zeta_{\ell}\perp\left(\sum_{f=1}^F\left(1 - \frac{\tau_f}{\beta_{2,f}}\right)\frac{W_f}{\log(2)}\left(\frac{\mat{\alpha}_{\mathcal{K},\ell}^{f,T}}{W_f\sigma^2 +{\mat{\alpha}_{\mathcal{K},\ell}^{f,T} \overline{\mat{p}}}}  - \frac{\mat{\alpha}_{\mathcal{K},\overline{\ell}}^{f,T}}{W_f\sigma^2 + \mat{\alpha}_{\mathcal{K},\overline{\ell}}^{f,T}\overline{\mat{p}}}\right) -R_{\ell}\right) \geq 0,\forall \ell \label{eq:KKTb} \\
&0\leq \theta_{k}\perp \left(P_{BS,k}^{max} - \mat{t}_{BS,k}^T\overline{\mat{p}}\right) \geq 0,\forall k \label{eq:KKTc} \\
&\overline{\mat{p}} \geq \mat{0} \label{eq:KKTc}
\end{align}
\end{subequations}
where $\zeta_{\ell}, \forall \ell \in \mathcal{L}$ and $\theta_{k},\forall k \in \mathcal{K}$ are the Lagrangian multipliers. Observe that the KKT conditions \eqref{eq:KKTa}-\eqref{eq:KKTc} are exactly same as the KKT conditions of Problem  \eqref{eq:P21}. Therefore, it implies that $\overline{\mat{p}}$ with the associated Lagrangian multipliers $\{\zeta_{\ell}, \theta_k\}$ is a KKT stationary solution to the original problem \eqref{eq:P21}. 
\end{IEEEproof}

\bibliographystyle{IEEEbib}
\bibliography{thesis2}
\ifCLASSOPTIONcaptionsoff
  \newpage
\fi

\end{document}